\documentclass{article}

\usepackage{arxiv}
\usepackage{subcaption}
\usepackage[utf8]{inputenc} % allow utf-8 input
\usepackage[T1]{fontenc}    % use 8-bit T1 fonts
\usepackage{hyperref}       % hyperlinks
\usepackage{url}            % simple URL typesetting
\usepackage{booktabs}       % professional-quality tables
\usepackage{amsfonts}       % blackboard math symbols
\usepackage{nicefrac}       % compact symbols for 1/2, etc.
\usepackage{microtype}      % microtypography
\usepackage{lipsum}		% Can be removed after putting your text content
\usepackage{graphicx}
\usepackage{doi}

\title{Some Pragmatic Prevention's Guidelines regarding SARS-CoV-2 and COVID-19 in Latin-America inspired by mixed Machine Learning Techniques and Artificial Mathematical Intelligence. Case Study: Colombia}
\author{Danny A. J. G\'omez-Ram\'irez \\ Instituci\'on Universitaria Pascual Bravo \\ Instituci\'on Universitaria ITM  (Instituto Tecnol\'ogico Metropolitano) \\ Visi\'on Real Cognitiva S.A.S.\\ Medell\'in, Colombia \\ \texttt{daj.gomezramirez@gmail.com}\\
\And Yoe A. Herrera-Jaramillo \\ Tecnol\'ogico de Antioquia - Instituci\'on Universitaria \\ Medell\'in, Colombia \\ \texttt{yoe.herrera@tdea.edu.co} \\
\And Johana C. Ortega-Giraldo\\ Visi\'on Real Cognitiva S.A.S.\\ Itagü\'i, Colombia \\ \texttt{jcog2583@outlook.com} \\
\And Alex M. Ardila-Garcia \\ Visi\'on Real Cognitiva S.A.S.\\ Medell\'in, Colombia \\ \texttt{Alex.Ardila.Garcia@gmail.com}}

\renewcommand{\shorttitle}{Some Prevention's Guidelines for COVID-19 based on Mixed AI techniques}

\usepackage{natbib}
\begin{document}

\maketitle

\begin{abstract}
We use an enhanced methodology combining specific forms of AI techniques, opinion mining and artificial mathematical intelligence (AMI), with public data on the spread of the coronavirus SARS-CoV-2 and the incidence of COVID-19 disease in Colombia during the first three months since the first reported positive case. The results obtained, together with conceptual tools coming from the global taxonomy of fundamental cognitive mechanisms emerging in AMI and with suitable contextual information from Colombian public health and mainstream social media, allowed us to stating specific preventive guidelines for a better restructuring of initial safe and stable life conditions in Colombia, and in an extended manner in similar Latin American Countries. More specifically, we describe three major guidelines: 1) regular creative visualization and effective planning, 2) the continuous use of constructive linguistic frameworks, and 3) frequent and moderate use of kinesthetic routines. They should be understood as effective tools from a cognitive and behavioural perspective, rather than from a biological one. Even more, the first two guidelines should be acknowledged in integral cooperation with the third one regarding the global effect of COVID-19 in human beings as a whole, this includes the mind and body.
\end{abstract}

	\section{Introduction}
	
	Each country in the world, without exception, faces an outstanding challenge regarding the generation of biological, medical, and chemically based therapeutic guidelines and solutions for overcoming the effects of COVID-19 disease. But also, each country has the challenge of developing effective policies and behavioural actions for minimizing, preventing, and overcoming all the negative effects of the SARS-CoV-2 virus, COVID-19's causing pathogen.
	In order to solve the first challenge, a wide plethora of techniques from artificial intelligence (AI), machine learning, computational biology, bio-medicine, and public policy initiatives have shown to be very valuable  \cite{bullock2020mapping}, \cite{alimadadi2020artificial},  \cite{vaishya2020artificial}.
	
	For the case of the second challenge, a combination of AI-based techniques such as opinion mining, or more specifically sentiment analysis \cite{hand2014data}, and the new emerging multidisciplinary paradigm of artificial mathematical intelligence  \cite{AMI} have a huge potential for providing contextually-robust and cognitively-sound pragmatic behavioural guidelines.
	
	It is worth noting that most of the existing specialized literature involving COVID-19 is mainly focused on biological, medical, and chemically based therapeutic guidelines, and significantly less efforts are focused on developing effective policies and behavioural actions \cite{bullock2020mapping}. In order to illustrate how AI-based techniques can be applied, we will focus our attention in the Latin American context, in particular, in the Colombian society.
	
	Since the behavioural patterns are structurally conditioned not only by the whole taxonomy of cognitive abilities of the mind (seen as both an individual and a collective), but also by the entire anthropological, cultural, environmental conditions (e.g. geographical), it is necessary to integrate a kind of minimal cognitive-anthropological blueprint of the Colombian population. Such a special kind of blueprint should be structured from a global proto-typical perspective in order to enhance the range of effectiveness that such (attitudinal) guidelines should possess.
	
	For the purposes of the present work, we will emphasize mainly the interaction component of such a blueprint due to the fact that this is one of the most relevant features for the spreading of SARS-CoV-2. In \S\ref{context}, we will provide and introduction to the origin and spread of SARS-CoV-2 in the world and its arrival and impact in Latin America and Colombia.
	
	In \S\ref{methods}, we will briefly introduce the methods used in our work, in order to give a more solid background to non-experts. This is very important in our current situation, where the whole (non-specialized) community needs access to more specialized information as self-contained as possible, and even more, when research work contains valuable results of preventive nature.
	
	In \S\ref{datamining}, we give a description of the main qualitative features of the information analysed (e.g. tweets of main trends in the Colombian context).
	
	Furthermore, in \S\ref{guidelines} we present the global landscape of findings, together with the taxonomy of main (behavioural) guidelines inspired partially by the Colombian social environment, but with a potential broader scope due to their ontological and universal nature. Such cognitive directions have a great potential to be applied to other countries with a relatively similar (anthropological) blueprint to Colombia.
	
	Lastly, in \S\ref{conclusions} we depict the major conclusions of our work and its potential extensions to the Latin American community, and potentially beyond.

	\section{Understanding the Context}
	\label{context}
	
	\subsection{COVID-19: Origin}
	
	The coronavirus disease of 2019 (COVID-19) is a multisystemic disease in humans that is caused by the beta-coronavirus SARS-CoV-2 (Severe Acute Respiratory Syndrome Coronavirus 2)  \cite{shereen2020covid}. SARS-CoV-2 is a zoonotic RNA virus that originated from wild bats  \cite{hu2015bat}. SARS-CoV2 is thought to have been transmitted to humans by pangolins, an intermediate host  \cite{ye2020zoonotic}, that are frequently traded for its scales in the black markets of Asia.
	
	The first outbreak of SARS-CoV-2 occurred in the city of Wuhan (China) and was reported on December 12th, 2019  \cite{guo2020origin}. There were 27 cases confirmed in the first report, but that escalated rapidly as over 84,000 people got infected in China in a period of six months  \cite{guo2020origin} (Table 1). The first confirmed case of SARS-CoV-2 infection outside China occurred in Thailand on January 13, 2020. The World Health Organization (WHO) declared a pandemic on March 13, 2020 due to the rapid propagation of SARS-CoV-2 in every continent. As of June 29, 2020, there were over 10.2 million confirmed SARS-CoV-2 infections and over half a million confirmed deaths in the world.
	
	\subsection{COVID-19 in Europe and the United States}
	
	Italy and Spain were the first two major outbreaks in Europe and the world outside of China. As of June 29, 2020, Italy and Spain were in the world's top ten in the total number of confirmed cases, deaths, and recovered individuals. In Spain, there were 248,970 confirmed cases, 28,346 deaths, and 150,376 recovered individuals by June 29, 2020 (Table \ref{tab:tabla_covid}). In Italy, there were 240,436 confirmed cases, 34,744 deaths, and 189,196 recovered individuals by the same date (Table \ref{tab:tabla_covid}). The outbreaks in Spain and Italy slowed down significantly in the month of June 2020 after strict lock-down and quarantine measures were imposed in the months of March, April, and May that resulted in significant death-toll reductions.
	
	\begin{table}[!htb]
		\begin{center}
			\begin{tabular}{lrrrr}
				\toprule
				Country        & Confirmed & Recovered & Deaths  & Deaths/100,000 people \\
				\midrule
				Colombia*      & 91,769    & 38.280    & 3,106   & 6.56             \\
				Brazil         & 1,344,143 & 746,018   & 57,622  & 27.43            \\
				Mexico         & 216,852   & 164,646   & 26,648  & 21.11            \\
				Peru           & 279,419   & 167,998   & 9,317   & 29.11            \\
				Chile          & 271,982   & 232,210   & 5,509   & 28.99            \\
				United States  & 2,562,921 & 685,164   & 125,928 & 38.37            \\
				Spain          & 248,970   & 150,376   & 28,346  & 60.39            \\
				Italy          & 240,436   & 189,196   & 34,744  & 57.30            \\
				United Kingdom & 313,467   & 1,364     & 43,659  & 65.50            \\
				Germany        & 194,910   & 177,812   & 8,976   & 10.81            \\
				Austria        & 17,723    & 16,420    & 703     & 7.94             \\
				China          & 84,757    & 79,610    & 4,641   & 0.33             \\
				
				\bottomrule
			\end{tabular}
		\end{center}
		\caption{COVID-19 confirmed reports in Colombia, The United Sates, China, and a selection of countries from Latin America and Europe. Sources: The John Hopkins University: Coronavirus Resource Center. June 29, 2020. Online: https://coronavirus.jhu.edu/map.html. *Instituto Nacional de Salud: Coronavirus en Colombia. June 29, 2020. Online: https://www.ins.gov.co/Noticias/Paginas/Coronavirus.aspx.}
		\label{tab:tabla_covid}
	\end{table}
	
	In the United Kingom, COVID-19 has also been devastating. There were 313,467 confirmed cases, 43,659 deaths (65.50 deaths/100,000 people), and just 1,364 official reports of recovered individuals as of June 29, 2020 (Table \ref{tab:tabla_covid}). By this date, The United Kingdom was the most affected region in Europe while Germany and Austria were at the opposing end when comparing death-tolls. In Germany, there were 194,910 confirmed cases, 8,976 deaths (10.81 deaths/100,000 people), and 177,812 recovered individuals (Table \ref{tab:tabla_covid}). In Austria, most of the reported cases (17,723) had recovered (16,420) and the death-toll was 703 (7.94 deaths/100,000 people)(Table \ref{tab:tabla_covid}) by June 29, 2020.
	
	The United States was the country most affected in the world by COVID-19 as of June 29, 2020. The reports up to then included over 2.5 million confirmed cases, over 125,928 confirmed deaths (38.37 deaths/100,000 people), and 685,164 recovered individuals (Table \ref{tab:tabla_covid}).
	
	\subsection{COVID-19 in Latin America: Arrival, Propagation, and Mortality}
	
	SARS-CoV-2 arrived in Latin American in February and March 2020 and it expanded rapidly in most countries  \cite{simbana2020interim}. It was brought to most countries by multiple infected individuals traveling from Europe (usually Spain, Germany, and Italy) and from the United States  \cite{rodriguez2020covid},  \cite{Laiton-Donato}. The first confirmed case of COVID-19 in Latin America was reported in Brazil on February 21, 2020  \cite{rodriguez2020covid}. This was two months after the initial outbreak in Wuhan, China. The first confirmed infected Brazilian national did not get infected in China  \cite{rodriguez2020covid}. Instead, this individual got infected with SARS-CoV-2 in Italy, which at the time had the largest outbreak outside China. As of June 29, 2020, Brazil was the second most affected country in the world by COVID-19 with over 1.3 million confirmed cases, 57,622 deaths (27.43 deaths/100,000 people), and 746,018 recovered individuals (Table \ref{tab:tabla_covid}).
	
	Peru, Mexico, and Chile were the most affected countries in Latin America after Brazil as of June 29, 2020. Peru was second in Latin America in the total number of confirmed cases (279,419), of which 9,317 died (29.11 deaths/100,000 people) and 167,998 recovered (Table \ref{tab:tabla_covid}). Mexico had the greatest number of confirmed deaths (26,648; 21.11 deaths/100,000 people) in Latin America after Brazil by June 29, 2020 (Table \ref{tab:tabla_covid}). The number of confirmed and recovered cases in Mexico until June 29 2020 was 216,852 and 164,646 respectively. In Chile, there were 271,982 confirmed cases that included 5,509 deaths (28.99 deaths/100,000 people) and 232,210 recovered individuals (Table \ref{tab:tabla_covid}). It is important to note that Brazil, Mexico, and Chile differ greatly in population size, but shared a similar approach to confronting COVID-19. These countries did not impose strict nation-wide lock-downs in the initial stages of their COVID-19 outbreaks.
	
	The propagation and daily mortality rates of the COVID-19 outbreaks accelerated in June 2020 in the majority of Latin American countries. This occurred after easing strict curfew preventive measures. These measures were imposed for one to two months (April and May) by the vast majority of countries in the region after confirming their first SARS-CoV-2 cases.
	
	\subsection{COVID-19 in Colombia}
	
\begin{figure}[!htb]
		\centering
		\includegraphics[width=0.6\textwidth]{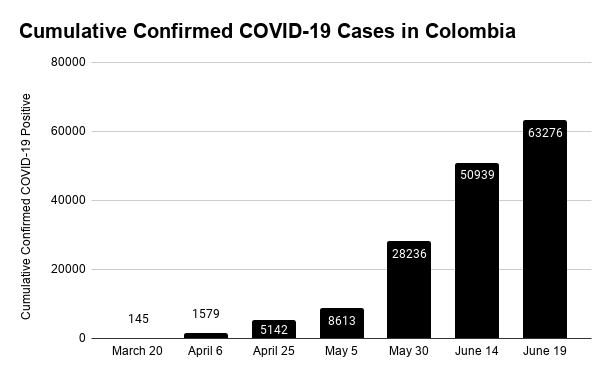}
		\includegraphics[width=0.6\textwidth]{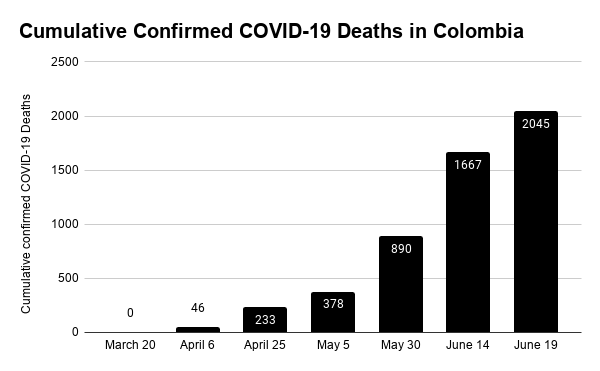}
		\includegraphics[width=0.6\textwidth]{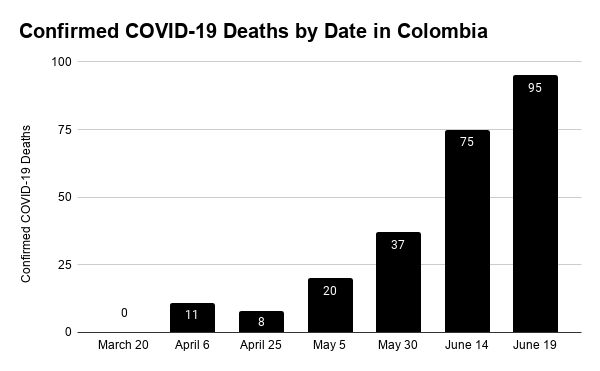}
		\caption{COVID-19 Colombia from March 6th to June 19th 2020. Source: Instituto Nacional de Salud: Coronavirus en Colombia. June 19, 2020. Online: https://www.ins.gov.co/Noticias/Paginas/Coronavirus.aspx.}
		\label{fig:cov1}
	\end{figure}

	The first confirmed case of COVID-19 in Colombia was reported on March 6, 2020 in Bogota, the capital and largest city of Colombia  \cite{Laiton-Donato}. The affected individual was a Colombian 19-year-old female that had returned to Colombia from Milan, Italy. By March 20, 2020, there were 145 confirmed COVID-19 cases in Colombia and zero confirmed deaths  (Fig. \ref{fig:cov1}). The first confirmed COVID-19 death in Colombia was reported on March 21st, 2020. This individual was a 58-year-old male who was a taxi driver in the touristic city of Cartagena. It is suspected that the taxi driver got infected after driving two Italian tourists on March 4. The taxi driver presented the first COVID-19 symptoms on March 6, the same day that the first confirmed COVID-19 positive was confirmed in Bogota, Colombia.
	
	The number of confirmed COVID-19 cases and deaths has increased exponentially in Colombia since March 6 (Fig. \ref{fig:cov1}). There were 1,579 confirmed cases and 46 deaths by April 6 (Fig. \ref{fig:cov1}).  These more than tripled by April 25, reaching 5,144 confirmed cases and 233 deaths (Fig. \ref{fig:cov1}). In May 2020, the spread of SARS-CoV-2 continued to increase rapidly. There were 8613 confirmed COVID-19 cases and 378 confirmed deaths on May 5. The figures nearly quadrupled by May 30 reaching 28,236 confirmed cases and 890 confirmed deaths (Fig. \ref{fig:cov1}). These figures doubled by June 14 with 50,939 confirmed cases and 1,667 confirmed COVID-19 deaths (Fig. \ref{fig:cov1}). By June 19, 2020, there were 63,276 confirmed COVID-19 cases and a cumulative total of 2,045 deaths. As of June 29, 2020, there were 91,769 confirmed cases, 3,106 deaths (6.56 deaths/100,000 people), and 38,280 recovered individuals (Table \ref{tab:tabla_covid}).

	The Colombian data reported here has been widely available to the Colombian public since the start of the COVID-19 outbreak. The Colombian president reported daily on COVID-19 in an one-hour long live television show that was broadcast
	live during prime-time by nearly all national television channels, radio stations, and official online platforms. In these reports, the president reinforced guidelines to prevent the spread of the SARS-CoV-2 virus that included mandatory use of face-masks in public, a two-meter distance from other individuals, and frequent hand-washing with soap and water. Official reports were open-access and published online daily by the National Institute of Health (Instituto Nacional de Salud). Official COVID-19 statistics and COVID-19 prevention guidelines were also reported daily by nearly all media on television, radio news, and newspapers.

	\section{Fundamental Methods}
	\label{methods}
	
	\subsection{Opinion mining}  \label{datamining}
	
	In order to understand the impact of COVID-19 preventive quarantine measures on Colombia's population, we carried out a task of sentiment analysis on one of the most popular platforms of massive use, Twitter. We chose seven important dates to analyze Colombians' reactions by measuring the polarity of their tweets (positive, negative, or neutral)  \cite{sail19}. Sentiment analysis is a branch of opinion mining which is a popular area of research in which the major goal is to find important information on texts or documents, more precisely, as stated by \cite{BL08}, opinion mining deals with the computational treatment of opinion, sentiment, and subjectivity in text (in alphabetical order).
	
	\subsubsection{Data collection}
	
	The tweets selected for this work were retrieved from the Twitter platform via a twitter API in Python. The selection criteria was that the tweets appear in one of the queries of the top ten trends on the selected dates at 8:00 p.m. EST according to the website: https://getdaytrends.com/es/colombia/.
	
	Table \ref{tab:dates_tw} summarizes the selected dates, the number of tweets collected, and the event that led to selecting each date for the present work.
	
	\begin{table}[!htb]
		\begin{center}
			\begin{tabular}{lcl}
				\toprule
				Dates        & Tweets & \hspace{2.5cm}Criteria                              \\
				\midrule
				March 20-21  & 583    & Colombian government announces mandatory COVID-19 preventive lockdown      \\
				&        & starting on March 24 through April 13               \\
				April 6-7    & 350    & First quarantine extension                       \\
				April 20-21  & 496    & Second quarantine extension                      \\
				May 5-6      & 276    & Third quarantine extension                       \\
				May 30-31    & 272    & Decentralization of COVID-19 policy, partial regional policy independence  \\
				June 14-15   & 304    & Number of daily cases surpasses 100 \\
				June 19-20   & 687    & First Tax-Free day                                      \\
				\midrule
				Total tweets & 2968   &                                                     \\
				\bottomrule
			\end{tabular}
		\end{center}
		\caption{Collection of tweets.}
		\label{tab:dates_tw}
	\end{table}
	
	\subsubsection{Data pre-processing}
	
	A tweet is a concise text containing opinions expressed in many different ways by different users. In November 2017, tweet length was extended to a maximum of 280 characters. A tweet can also contain abbreviations and symbols as emojis. For these reasons, the raw data obtained in these cases are highly susceptible of inconsistency and redundancy, and hence, pre-processing is crucial. This is and should be the first step on any twitter analysis.
	
	In this work, pre-processing is performed in Python with the frameworks of Natural Language Package ToolKit (NLPTK) and sklearn. We proceeded in the following order:
	
	\begin{itemize}
		\item Remove all URLs, hashtags (\#topic) and targets (@username).
		\item Translate the tweets gathered with googletrans module from Spanish to English.
		\item Remove punctuation marks, symbols, numbers.
		\item Tokenization, that is, separate tweets in its constituing words.
		\item Common stop words are removed, e.g., what, a, the, of, \ldots
		\item Convert all text to lower case to make the dataset uniform.
		\item Delete repeated characters, e.g. sweeeet was replaced sweet.
		
	\end{itemize}
	The set of all unrepeated preprocessed words constitute the so-called "Corpus" or vocabulary. This corpus is the base for the sentiment analysis and its the real purpose  of  the data pre-processing process. It is important to notice that we are working with seven (7) datasets of tweets, so pre-processing and corpus were obtained for each individual of dataset.
	
	\subsubsection{Feature extraction}
	
	The following step in the sentiment analysis is the extraction of features from the tweets with respect to the corpus corresponding to each dataset. Let $C=\{w_i \}_{i=1}^N $ be the corpus of tweets of a given date, that is, $C$ contains all the words in the tweets from the given date with no repetition. The vectorized tweet $vect(T_i)$ assigned to tweet $T_i$ is defined by $vect(T_i) = [v_k]_{k=1}^N
	$, where $v_k$ is the number of times the word $w_k$ appears in tweet $T_i$, we do this by following \cite{lee19}.
	The function we used to carry out this process of vectorization in Python is $Countervector$ from the Natural Lenguage ToolKit (NLTK) module. As an illustration of this process, consider $T_1$ the first tweet of March 20 and 21, which reads ``authorities wait declare mandatory quarantine Colombia", then the first six (6) entries of $vect(T_1)$ are 1's and the rest are zeroes. The second pre-processed tweet is $T_2$: ``control measures Italy Colombia trapped Italy outside Colombia", then $vect(T_2)$ has its first five (5) entries zero and its sixth entry is 2 because it contains the word Colombia twice and Colombia was the six word in the first tweet, the entries 7, 8, 10, 11 are 1 because they correspond to the words control, measures, trapped and outside, and the entry 9 is 2 because Italy appears twice. These examples are illustrated in Table \ref{tab:vect_tw}.
	
	\begin{table}[!htb]
		\begin{center}
			\begin{tabular}{lc}
				\toprule
				Tweet   $T_i$                        & Vectorized tweet $vect(T_i)$                               \\
				\midrule
				``authorities wait declare mandatory & $[1\; 1\; 1\; 1\;1\; 1\; 0\; 0\; 0\;0\;0\;0\;\cdots \; 0]$ \\
				quarantine colombia"                 &                                                            \\
				``control measures italy colombia    & $[0\; 0\;0\; 0\;0\; 2 \; 1\; 1\; 2\;1\;1\;0\;\cdots\; 0]$  \\
				trapped italy outside colombia"      &                                                            \\
				\hspace{2cm}\vdots                   & \vdots                                                \\
				\bottomrule
			\end{tabular}
		\end{center}
		\caption{Vectorization of tweets.}
		\label{tab:vect_tw}
	\end{table}
	
	In this way, we get a matrix known as the term document matrix (TDM). In TDM, rows correspond to the words from the corpus and the columns correspond to the tweets. Hence, the size of a TDM corresponds to the pair $(N,m)$ where $N$ is the number of words after pre-processing and $m$ is the number of tweets given in Table \ref{tab:dates_tw}. For each of the datasets considered, the number $N$ obtained was $2914$ for March 20-21, $2145$ for April 6-7, $2929$ for April 20-21, $1968$ for May 5-6, $1951$ for May 30-31, $2073$ for June 14-15, and $2986$ for June 19-20.
	
	\subsubsection{Sentiment analysis}
	
    Our goal was to determine the polarity and sensitivity of each tweet in our datasets. For computing polarity and sensitivity of a tweet we make use of the TextBlob module in Python. TextBlob is a python library and offers a simple lexicon-based API to access its methods and perform basic Natural Language Programming (NLP) tasks, including sentiment analysis. Polarity of a word is a real number between -1 and 1, where a negative value corresponds to a negative word, 0 to a neutral word, and a positive value to a positive word. The polarity of a sentence or document is then calculated by the average of the polarity of its constituting words. Subjectivity on the other hand, is a value from 0 to 1, where 0 is assigned to the word if it is a fact. The higher the subjectivity the more opinionated the word is. The subjectivity of a sentence or document is then defined as the average of the subjectivity of its words, \cite{uss18}.
	
	We take advantage of the TDMs to extract the words with more frequency on each of the tweet datasets. This was performed by adding the column values, which contains the number of times the corresponding word appears in each tweet. By extracting the most common words, we can deduce the most common topics. Nevertheless, we preferred to visualize the most common topics by using the wordcloud module in Python to represent these topics for each of dates chosen.

	\subsubsection{Analysis of results}\label{sec:results}
	Here we show the graphical results and conclusions of our research in regards to the sentiments perceived in  tweets obtained from Colombia's inhabitants during the COVID-19 pandemic. First, we analyze in detail the first and the fourth datasets because of the results obtained. Second, for the most interested reader we analyze in a general fashion the rest of the datasets  shown in Appendix \ref{ap:ap} .  

\begin{figure}[!htb]
		\centering
		% include first image
		\includegraphics[width = 8cm]{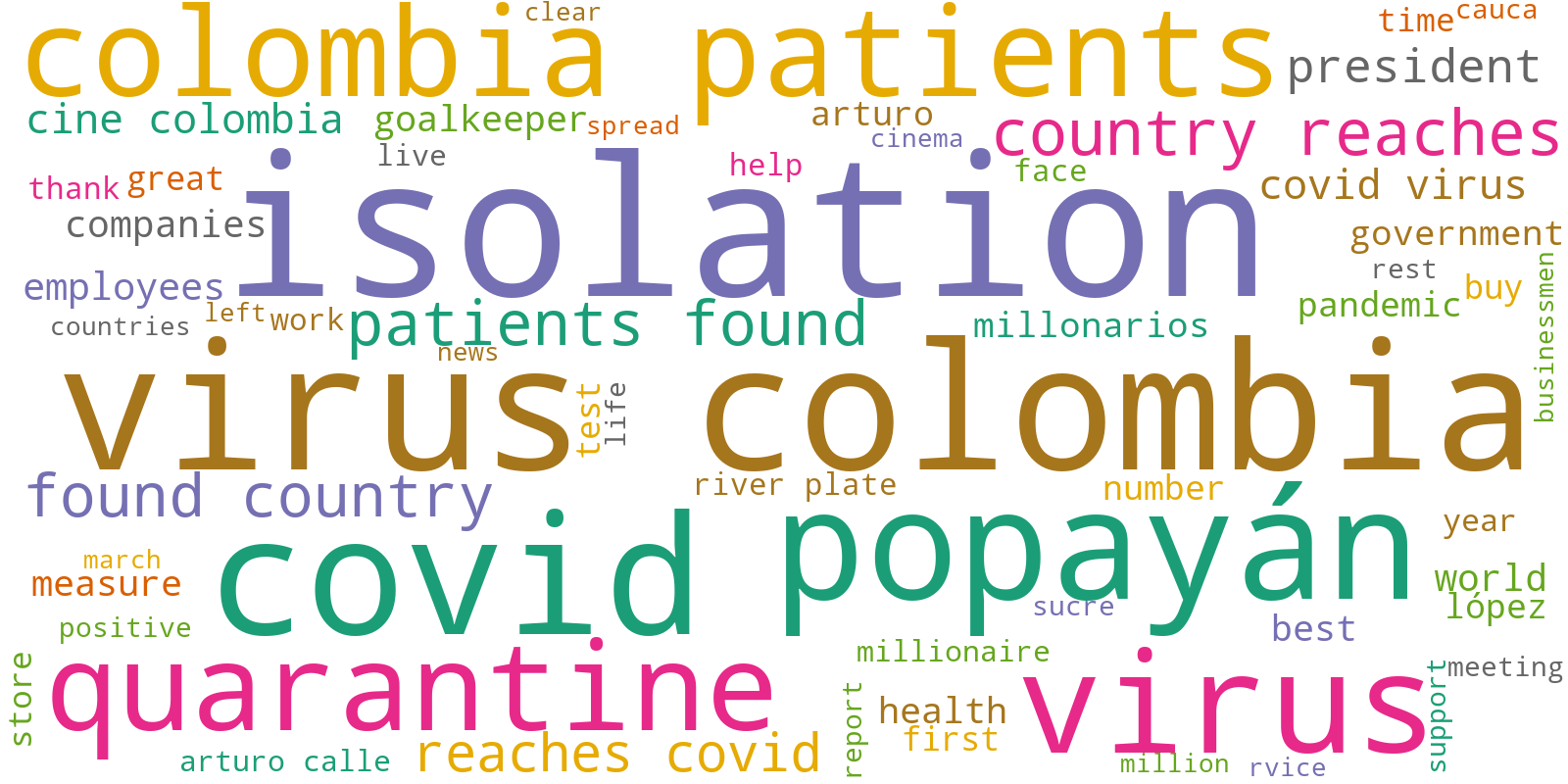}
		\caption{Wordcloud of March 20 and 21}
		\label{fig:day1_wc}
	\end{figure}

From the tweet dataset from March 20 and 21, we can highlight from the wordcloud (Fig. \ref{fig:day1_wc}) that the most common topics were related to COVID-19 and the preventive quarantine decree ordered by the Colombian government to slow down the spread of COVID-19. This dataset also shows the concern among Colombians about the growing number of confirmed cases in the country.
	
	%%%%%%%%%%%55
	%%%%%%%%%%%%%%%%55
	%%%%%%%%%%%%%%%%%%5
\begin{figure}
		\begin{subfigure}{0.4\linewidth}
			\centering
			% include second image
			\includegraphics[height=4cm,width=5cm]{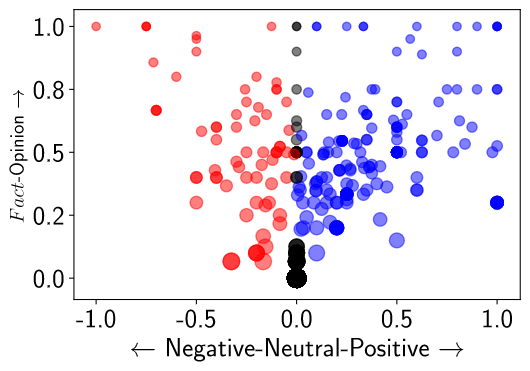}
		\end{subfigure} \hspace{0.2cm}
		\begin{subfigure}{0.5\linewidth}
			\centering
			% include third image
			\includegraphics[height=4cm,width = 7cm]{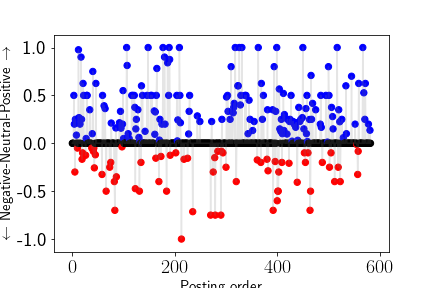}
		\end{subfigure}
		\caption{Polarity and subjectivity of tweets on March 20 and 21.}
		\label{fig:day1_polsub}
	\end{figure}

	In Fig. \ref{fig:day1_polsub} (left side), we can observe that there were more positive tweets (blue dots) than negative ones (red dots). This is also displayed in detailed in Table \ref{tab:tab_summary}, which summarizes the number of positive, neutral, and negative tweets by date. On March 20 and 21, the more prevalent type of tweets were neutral. We can also note in Fig. \ref{fig:day1_polsub} (right side) that positive and negative tweet posting orders were equally distributed with respect to the posting order.

	\begin{table}[!htb]
		\begin{center}
			\begin{tabular}{lrrrr}
				\toprule
				Dates       & Positive & Negative & Neutral & Total \\
				\midrule
				March 20-21 & 169      & 68       & 346     & 583   \\
				April 6-7   & 167      & 56       & 127     & 350   \\
				April 20-21 & 136      & 62       & 298     & 496   \\
				May 5-6     & 76       & 112      & 88      & 276   \\
				May 30-31   & 98       & 46       & 128     & 272   \\
				June 14-15  & 115      & 61       & 128     & 304   \\
				June 19-20  & 187      & 98       & 402     & 687   \\
				Total       & 948      & 503      & 1517    & 2968  \\
				\bottomrule
			\end{tabular}
		\end{center}
		\caption{Summary of sentiment analysis.}
		\label{tab:tab_summary}
	\end{table}

\begin{figure}[!htb]
		\centering
		% include first image
		\includegraphics[width = 8cm]{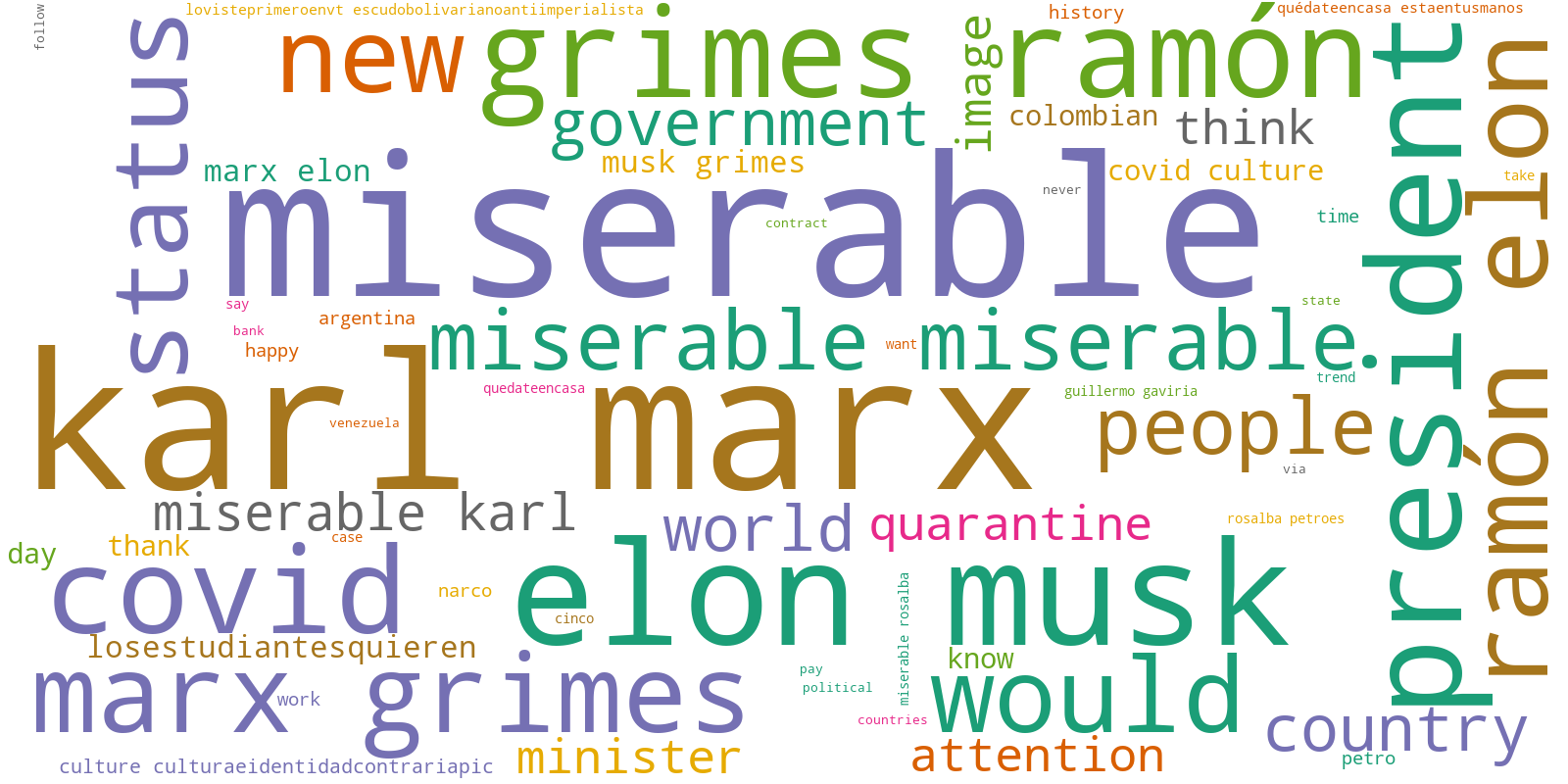}
		\caption{Wordcloud of tweets on May 5 and 6.}
		\label{fig:day4_wc}
	\end{figure}

	As it can be appreciated in Table \ref{tab:tab_summary}, neutral tweets prevailed in all dates selected, except on May 5 and May 6. On these dates, there were 112 negative tweets of the 276 analyzed. That is, 40\% of the tweets in this dataset were from people writing on negative feelings. It is worth noticing that a third extension on the mandatory isolation was announced on May 5 and that the main topics in twitter were miserable, Karl Marx, COVID, and grimes (Fig. \ref{fig:day4_wc}).
	%%%%%%%%%%%%%%%%%%%5
	%%%%%%%%%%%%%%%%%55
	%%%%%%%%%%%%%%%55
	
\begin{figure}
		\begin{subfigure}{0.4\linewidth}
			\centering
			% include second image
			\includegraphics[height=4cm,width=5cm]{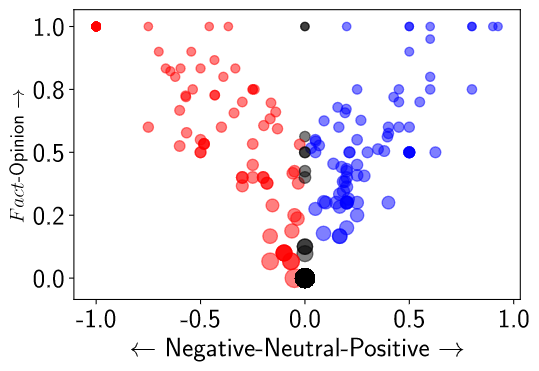}
		\end{subfigure}
		\begin{subfigure}{0.5\linewidth}
			\centering
			% include third image
			\includegraphics[height=4cm,width = 7cm]{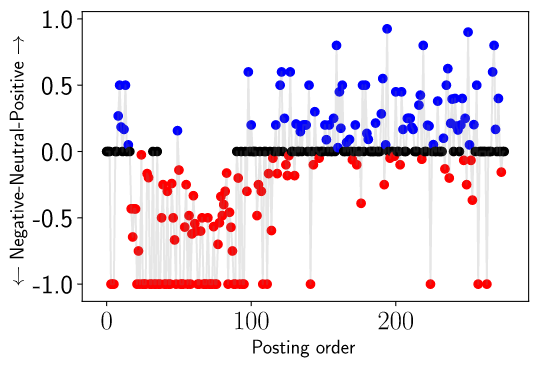}
		\end{subfigure}
		\caption{Polarity and subjectivity of tweets on May 5 and 6}
		\label{fig:day4_polsub}
	\end{figure}

	In addition, we notice that approximately Tweets 20 to 80 were negative as shown in Fig. \ref{fig:day4_polsub} (right) suggesting that one of the trends was very negative and strong because of their polarities is $-1$. An example of such tweets was tweet $T_{23}$, which read: “miserable playing health Colombia, squandering billions pesos armored cars, advertising wash image esmad midst pandemic." The subjectivity of this tweet  was 1, meaning that this tweet was absolutely an opinion. This tweet was referring to the Colombian government buying security cars for considerable amounts of money.	
	
Based on the data from the other dates examined, we can state that twitter users in Colombia during the first three months of the COVID-19 outbreak in Colombia were focused on topics other than the pandemic and quarantine. Colombians were centered on the pandemic and quarantine only on the first days (March 20 and 21) we analyzed. As time went by, it was noticeable that Colombians started to express neutral sentiments about the pandemic and that their focus switched to different topics such as corruption, the economy, and entertainment, especially soccer. This occurred in spite of the alarming rate of spread of the COVID-19 pandemic in Colombian and Latin America.

\begin{figure}[h]
		\begin{subfigure}{0.4\linewidth}
			% include second image
			\includegraphics[height=4cm,width = 5.2cm]{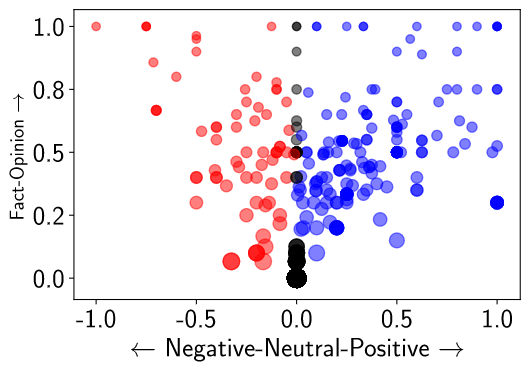}
		\end{subfigure}
		\begin{subfigure}{0.5\linewidth}
			\includegraphics[height=4cm,width = 7cm]{mar20_pol_ch.png}
		\end{subfigure}
		\caption{Sentiment analysis of April 6 and 7}
		\label{fig:day2}
	\end{figure}

By April 6 and 7, we can observe that even though the total number of confirmed COVID-19 cases was 1,579 and that 46 deaths were reported in total, people were not yet fully aware of the future effect of COVID-19 in their lives. This is shown in the worldcloud (\ref{fig:day2}) where it is evident that Colombians were concerned about topics other than the pandemic. On April 6 and 7, Colombian twitter users were most concerned about a political scandal involving the Colombian ambassador in Uruguay. They were also concerned about losing access to college education and jobs because of the pandemic. In terms of polarity, we can state that polarity was evenly distributed in posting order (\ref{fig:day2} bottom right). Forty eight percent of the tweets were neutral and the level of subjectivity indicated that the vast majority of tweets were opinions.

\begin{figure}[h]
		\begin{subfigure}{0.4\linewidth}
			\includegraphics[height=4cm,width = 5.2cm]{mar20_pol.png}
		\end{subfigure}\hspace{1cm}
		\begin{subfigure}{0.6\linewidth}
			\includegraphics[height=4cm,width = 7cm]{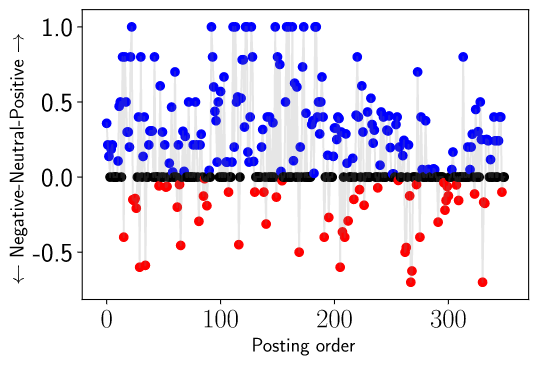}
		\end{subfigure}
		\caption{Sentiment analysis of April 20 and 21}
		\label{fig:day3}
	\end{figure}

For April 20 and 21, the wordcloud (\ref{fig:day3}) showed that the economy was the dominant twitter topic, especially oil prices and the dollar. Polarity in these tweets shows that 60\% of tweets were neutral, 13\% negative, and 27\% positive. The majority of tweets were neutral and positive even though the wordcloud showed that tweets appeared to express Colombians were very concerned about the economy. 	

\begin{figure}[h]
		\begin{subfigure}{0.4\linewidth}
			% include second image
			\includegraphics[height=3.6cm,width=5cm]{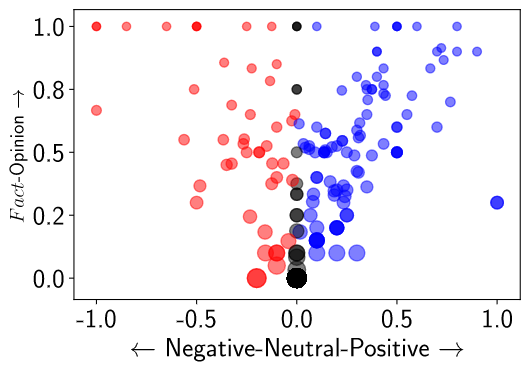}
		\end{subfigure}\hspace{1cm}
		\begin{subfigure}{0.5\linewidth}
			% include third image
			\includegraphics[height=4cm,width = 7cm]{mar20_pol_ch.png}
		\end{subfigure}
		\caption{Sentiment analysis of May 30 and 31}
		\label{fig:day5}
	\end{figure}

We observed that people were not talking much about the coronavirus on May 30 and 31 anymore. On these dates, the decentralization of COVID-19 pandemic's measures was decreed, each state acquired some autonomy to impose or liberate restrictions in their territories. The most trending topics on these dates were the SPACEX shuttle, tourism, and future challenges  (Fig. \ref{fig:day5}). There were only 17\% of negative tweets, but 36\% were positive and  47\% were neutral.

The main topics on Twitter in Colombia on June  14 and 15 were soccer, the Colombian soccer player James Rodriguez and his soccer club (Real Madrid). COVID-19 did not show up in the worldcloud even though there were already an increased number of confirmed COVID-19 cases (1579) and deaths (46). We can also see that by June 14 and 15 there were 20\% of negative tweets only and that the majority were neutral.  This is surprising because there were 75 COVID-19 deaths reported on June 14 alone and people appeared not to be talking about this issue.

\begin{figure}
		\begin{subfigure}{0.4\linewidth}
			% include second image
			\includegraphics[height=3.6cm,width=5cm]{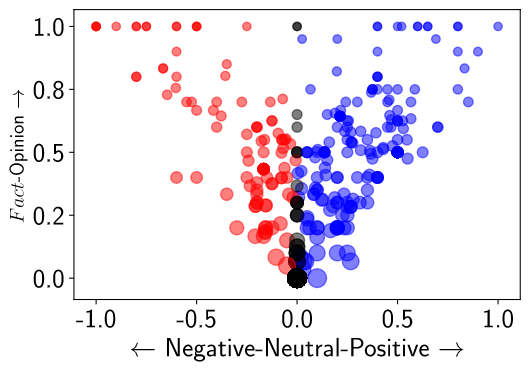}
		\end{subfigure}\hspace{1cm}
		\begin{subfigure}{0.5\linewidth}
			% include third image
			\includegraphics[height=4cm,width = 7cm]{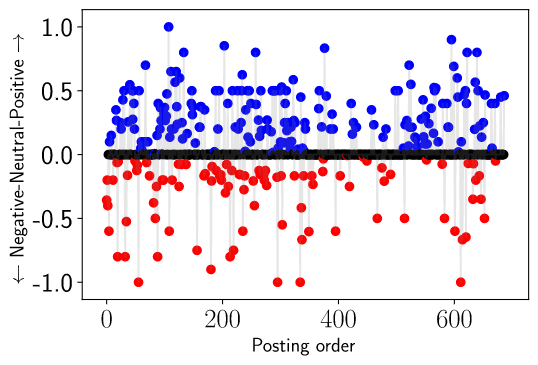}
		\end{subfigure}
		\caption{Sentiment analysis of June 19 and 20}
		\label{fig:day7}
	\end{figure}

The last dates (June 19 and 20) selected in the present study were the dates the Colombian government had declared a tax exemption in all purchases to reactivate the Colombian economy. The COVID-19 death-toll on June 19 was 95 deaths and a total of 2,045 deaths since the start of the COVID-19 outbreak in Colombia. These numbers did not appear to be of concern for most Twitter users in Colombia. Perhaps, most people were tired of being under lockdown and were more interested in other topics such as the tax-free day. This day became an opportunity and perhaps good reason for most people to leave their homes and go shopping. These resulted in large messy gatherings at shopping malls for which there were not sufficient preventive measures to avoid crowds. Quarantine and social distancing appeared to be something distant in the past. People were very excited and anxious about shopping and this was reflected on Twitter. However, in spite of this, the majority of tweets were neutral (51\%),  26\% were positive, and 13\% were negative (\ref{fig:day7}).

\subsubsection{Machine learning methods}
Supervised machine learning techniques are used for classifying problems. The classification algorithm learns from a training dataset with known categories for each training input and then predicts the category for an unseen input. The performance of the method is measured with a testing set, that is, a dataset of input with known categories but not seen by the machine in the training process. In this work we compare three (3) different sentiment classifiers of the supervised type with the dataset of tweets collected.

For evaluating the performance of a classifier, the confusion matrix $C=[c_{ij}]$ is used. 
The entry $c_{ij}$ is the number of testing items true class i that were classified as being in class j. The two metrics we used for measuring the performance of the classifier is obtained by taking the weighted average of the accuracies and precisions for each class or category as given by Formulae (\ref{eq:ac}) and (\ref{eq:pr}). 
\textit{Accuracy:} 
\begin{equation} \label{eq:ac}
Acurracy_i = \frac{TP_i+TN_i}{TP_i + TN_i + FP_i + FN_i}
\end{equation}
\textit{Precision:} 
\begin{equation} \label{eq:pr}
Precision_i = \frac{TP_i}{TF_i + FP_i}.
\end{equation}
Here $TP_i$ is the number of testing inputs correctly classified in the $i-th$ class; $FP_i$ is the number of inputs which are wrongly classified in the category $i$; $FN_i$ is the number of inputs from category $i$ that are misclassified and $TN_i$ is the sum of the remaining entries of the diagonal of the confusion matrix. 

The are two options for the weighted average: the Macro-level which consists of given the same weight for every class, and the Micro-level evaluation metrics, which gives each item equally, and making classes with more observations heavier than other with lesser items.

\textbf{Support vector machine (SVM)} \cite{leb15} This method consists of assigning a vector representation of the labeled training dataset as points in space, mapped such that the members of the separate categories are divided by a gap as wide as possible. The linear SVM separates these categories by linear subspaces. The testing dataset vectors are then mapped into the same space and predicted to belong to one of the categories based on which side of the gap they fall in.

In mathematical terms, a support vector machine provides a set of hyperplanes in an infinite dimensional space. The regions determined by these hyperplanes become the categories of the data. It is clear that the hyperplane which gives the best separation is the one that is farthest from all the nearest training examples (the support vectors) of each category. Further, the larger the margin, the lower the generalization error of the classifier will be.

\textbf{Naïve Bayes (NB)} \cite{leb15} This method generates a dataset with the frequency tables of the labeled data. With this dataset, a table with the corresponding probabilities of occurrences is created. The algorithm utilizes the Naive Bayes Equation (\ref{eq:nb}) for the probabilities. 
\begin{equation} \label{eq:nb}
P(c|d) = \frac{P(c)P(d| c)} {P(d)}.
\end{equation}

The category or class of an input corresponds to the highest probability obtained. Hence, 
the NB Algorithm assigns an input $d$ to the class $c$, that maximizes $P(c|d)$.

$k$\textbf{-Nearest neighborhood (K-NN)} \cite{sou18} It is a non-parametric algorithm, usually very effective in most of the cases. It is also one of the most popular for text categorization, data mining, and many more. The $k$-NN algorithm works by analogy, that is, compares the unknown data point with the training data points to which it is close as measured by Euclidean distance. Through the combination of a number
of local distance functions based on individual attributes, a global
distance function can be calculated. The distance is precisely the weighted of the point to the $k$ nearest neighbors. 

\textbf{Results.} For the implementation of these algorithms we use several configurations. The results of the ones with the best performance are presented in Table \ref{tab:mchres}. We split the  collection of all the tweets from the seven datasets into a training part (85\%) and a test part (the remaining 15\%. For the KNN method we use the linear algorithm $k = 3$ neighbors. As it can be appreciated the Linear SVM algorithm outperformed the others soundly.

\begin{table}
	\centering
	\begin{tabular}{lccc}
		\toprule
		Classifier &   Accuracy &  Precision micro &  Precision macro \\
		\midrule
		NB &  72.76 &        72.77 &        68.15 \\
		k-NN &  59.15 &        59.15 &        54.97 \\
		Linear SVM &  82.13 &        82.13 &        78.85 \\
		\bottomrule
	\end{tabular}
	\caption{Performance evaluation of the machine learning algorithms.}
	\label{tab:mchres}
\end{table}

	\subsection{Artificial Mathematical Intelligence}
	
	The research program of artificial mathematical intelligence (AMI) is a multidisciplinary meta-project with the aim of setting all the necessary theoretical and computational  foundations for constructing (general versions of) a Universal Mathematical Artificial Agent (UMAA), which essentially is able to serve as co-creative formal assistant for any kind of researcher who used mathematically-based frameworks in his/her research. Specifically, an UMAA will be able to help in the purely conceptual part of the frameworks, e.g. solving a (solvable) mathematical conjecture (i.e. finding a formal proof or giving a counterexample), giving conceptual hints in order to solve faster the conjecture. In other words, the AMI program envisions to create the most wide a powerful form of artificial intelligence using, combining and fusioning techniques and concepts of (computational) cognitive sciences, computational creativity, computational logic, pure and applied (meta)mathematics, theoretical physics, classic AI, computational linguistics and philosophy of mind, among others  \cite{AMI}.
	
The main three pillars of AMI are the following:
	firstly, the new cognitive foundations of mathematics\' program aiming to found new and renewed multidisciplinary foundations for (working) mathematics and for the way in which the mind generates mathematics from a pragmatic perspective.
	Secondly, the identification and meta-formalization of a global taxonomy of all the essential cognitive (metamathematical) abilities that the mind uses for generating (abstract) mathematical structures, conjectures and theories.
	Lastly, the implementation of all the former notions and structures into computational languages and software being able to simulate initial (and local) versions of a UMAA in a pragmatic and human style way.
	
	Due to the wide methodological spectrum of the second pillar of AMI, there are fundamental aspects of this meta-cognitive taxonomy that can be used not only for the original goals of AMI, but that can be applied for understanding and forecasting qualitatively behavioural patters of individual and collectives. In particular, in the context of this chapter we will use some of the seminal results of this taxonomy together with specific contextual information coming from data mining techniques for designing pragmatic guidelines for preventing (and overcoming) and potential infection with SARS-CoV-2 (and COVID-19) in the current Latin American context.

	\section{Pragmatic Preventive Guidelines}
	\label{guidelines}
	Based on the former qualitative and quantitative results obtained by understanding the (Latin American) context, by doing a relevant sentiment analysis via data mining applied upon a relevant collection of tweets, by using some of the most relevant results of the artificial mathematical intelligence meta-project concerning cognitive fundamental mechanisms, and finally by integrating addition behavioural results, we will present here some pragmatic performance guidelines for increasing the quality of life during (permanent) confinement and gradual return to a public normality. Our view of health is very holistic, meaning that a human being is integrated by a visible (e.g. physical) component and by an invisible (e.g. cognitive) one, and only in a suitable balance of sound conditions between both dimensions a genuine and compact health is truly obtained.
	Local governments and public health organizations should be cautious about advising people on how to enlighten and integrate both components of health and incorporate them into their policies.
	
	\subsection{Regular Creative Visualization and Effective Planning}
	
	It is a well-known fact in neuropsychology that when the human brain (sometimes referred as our subconscious mind) is exposed either to sensory information (e.g. images) or to purely phenomenological information (e.g. mental self-created images) cannot differentiate between reality and fantasy  \cite{gawain2016creative}.
	Moreover, creative visualization has a wide number of advantages into the cognitive health of human beings. One of the most important is that suitable techniques of visualization possess the power, the speeding up, the fulfillment of personal goals, and of giving a more authentic sensation of self-realization.  \cite[Part III-IV]{gawain2016creative}.
	Furthermore, in an environment where leaving home turns out to be so challenging and one should be long periods of time inside a building, creative visualization (e.g. involving desired empowering natural environments) would have the positive cognitive effect of giving a genuine sensation of being there (at least from a phenomenological perspective). Such a sensation has, in fact, a positive influence into the body since the neurological networking patterns responsible for processing the sensory-perceptual information coming from having visited such empowering natural environment in a life manner, and for having visited them phenomenologically are quite similar; and so, they affect the rest of the body in a resembled manner.

	Now, after using creative visualization for generating more clear and specific visions and goals in life, the following natural steps is to be able to create realistic (written) short-, medium- and long-term plans  for the achieving of these objectives. Effective and specialized planning is not only highly practical for companies and any kind of society, but also for individuals. Even more, in a context of a pandemic, to construct a subjective feeling and a perception of potential achievement of goals turns out to be very useful for the emotional states of the mind, since they replace an excessive focus of attention on the disadvantages of the pandemic and shift the cognitive emphasis on the advantages of the environment (for the fulfillment of the personal visions, for example).

	\subsection{Using Constructive Linguistic Frameworks}

	The particular language we use in our daily life, the words we choose to hear and to speak play a crucial role in the way we think. In particular, they play a seminal component in the way we create new ideas and in the way we combine them efficiently. In fact, a huge amount of the things that we do in our daily lives are previously configured in a syntactic manner and pronounced in a phonetic way in our mouths  \cite{thewaywethink}.
	In particular, one of the most fundamental cognitive-linguistic abilities that our mind uses for constructing a coherent subjective model of reality is metaphorical thinking  \cite{lakoff1980}. Roughly speaking, a metaphor is saying that a concept $X$ is (like) another concept $Y$. For example, "Life is (like) a Game" is a metaphor. Moreover, metaphorical reasoning is one of the most fundamental cognitive meta-mathematical abilities for generating formal mathematical knowledge, and in an extended manner for doing scientific research  \cite[Part II]{AMI}.
	
\begin{figure} 
		\centering
		\includegraphics[width =8cm]{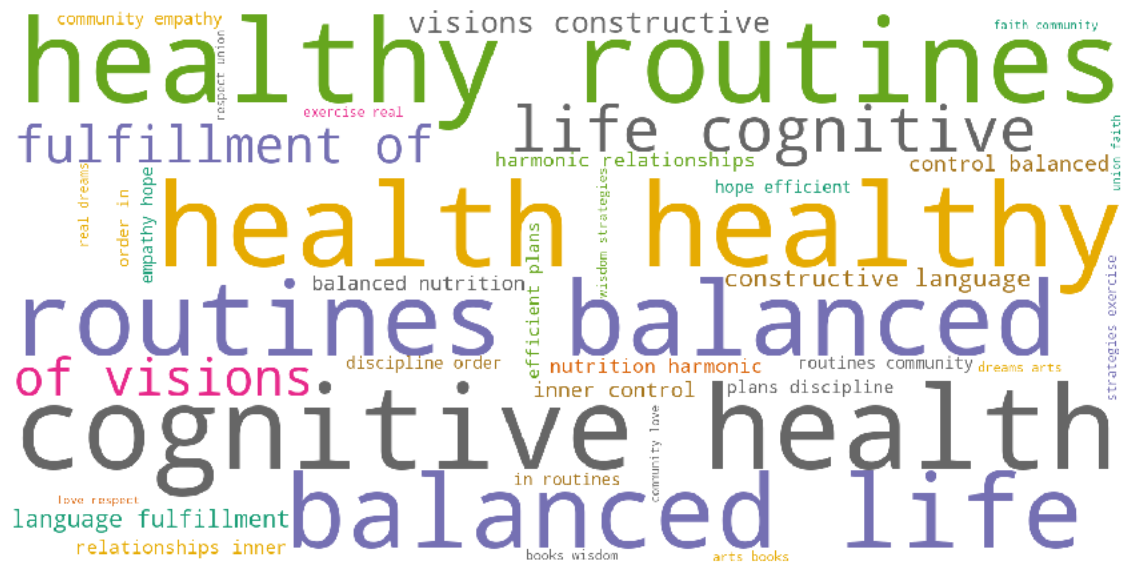}
		\caption{Desired wordcloud}
		\label{lastwordcloud}
	\end{figure}

	So, in the context of the COVID-19 pandemic it is fundamental to be able to build constructive, useful and inspiring metaphors for life, development, and personal success based on inspiring works instead of destructive ones. In our case, the section \ref{datamining} shows the morpho-syntactic evolution of hundreds of Colombian tweets for three months. From the evolution of the graphics one sees that the main trends in Colombia use gradually less and less words directly related with the COVID-19 and they started to evolve to other kind of topics like sports and space launches, politics, among others.
	This evolution into more constructive existential topics and notions is desirable, since an excessive exposition to destructive notions related with COVID-19 and with the increasing number of deaths could lead to premature depressive patterns and to an emotional atmosphere of global panic, which would deteriorate the confinement conditions and the inner harmony of the families. However, one can also appreciate in all the graphics shown in \S\ref{datamining} a tendency of amplifying issues of secondary importance related to the pandemic situation and not completely positive tendency of the feelings described in the majority of the tweets. Therefore, here we aim to decisively invite the usage of an explicit positive language aiming to renew the immediate social, mental and financial (subjective) conditions of each person. This would be an initial plus for being able to materialize the desired goals and visions that each person wishes to have in order to overcome all the challenges that the current outbreak has started. In particular, the formulation of empowering and inspiring metaphors of personal (and communal) use possess a huge force for boosting new and better habits for restructuring a more solid, stable and altruistic way of living (as individual and as a collective (e.g. family)).
	In the Fig. \ref{lastwordcloud}, one can see an ideal wordcloud containing the essential aspects described in the last two guidelines.

	Note that a similar linguistic behavioural approach can be applied virtually for any kind of cognitive (metamathematical) mechanism like analogical thinking, conceptual blending, conceptual substratum, exemplification, particularization, generalization, among others  \cite[Part II]{AMI}.
	
	\subsection{Frequent and Moderate Kinesthetic Routines}
	
	The beneficial and therapeutic effects of regular exercise routines for a healthy life are very well-known in medicine and psychology (see, for instance  \cite{johnson1985exercise},  \cite{ross2010health} and  \cite{sparkes2013}). Now, due to the fact that the SARS-CoV-2 is a virus that attacks and debilitates the respiratory system, it seems fundamental, from a preventive perspective, to integrate strengthening routines for the breathing system; so, one of the most natural and cheapest of them is without any doubt regular sporting activities. Even more, in a broader context beyond Latin America, two of the most common co-morbidity factors for people dying of COVID-19 are arterial hypertension (and diabetes)  \cite{yang2020prevalence}, \cite{richardson2020presenting}, which can also be prevented by doing exercise on a regular basis  \cite{ciolac2012}.
	Thus, our third prevention guideline is to do intelligent (indoor) exercise routines like continuous moderate-intensity training (CMT) routines with a duration or at least 30 minutes or more, 3-4 times a week. Here is important to create the routine first with small and pleasant sessions orchestrated with motivating and loved music pieces. Due to the spatial restrictions caused by the pandemic, it is highly recommended to have access to indoor running machines, stationary bicycles, stair-climbers or to following encouraging online exercise routines like aerobics, Zumba, salsa (or any kind of dance style), martial arts, among many others.
	
	The most challenging part requires to create the routine at the beginning, and the more you have repeated the sessions (moderately), the easier it is, the deeper it influences your unconscious mind and the greater of the benefit.

	\section{General Conclusions}
	\label{conclusions}
	Due to the wide spectrum of anthropological dimensions that the human being possesses, a genuine state of holistic health can be achieved taking care not only of the pure biological and physiological dimension of people (understood as individuals and as communities), but also taking really precise care of the cognitive dimension. Thus, based in a suitable combination of techniques coming from data mining, artificial mathematical intelligence and implicitly from cognitive sciences we offer simple and, at the same time, powerful behaviour techniques or guidelines for maintaining the mental health while preserving (longer) public policies of confinement due to the spread of COVID-19 in Colombia, and in other Latin American countries with similar contextual conditions. In fact, based on the universality of laws governing the functioning of the mind, it is worth to study the wide potential of applying the same kind of preventive guidelines to virtually any kind of society.\footnote{Assuming, of course, a suitable contextualization of each of them to the corresponding local policies and cultural conditions.}

	Specifically, we describe guiding principles relates with suitable visualization and planning techniques\footnote{These kind of techniques possess a wide and strong support from groups of people going beyond the academic field, like, for example, particular communities of entrepreneurs or athletes.}, the use of inspirational linguistic schemes and the generation of habits related with empowering kinesthetic practices. 
	
	In this chapter, we focus only on the former three main preventive guidelines. However, due to the wide spectrum of complexities of the human mind, a lot of additional guidelines could be analyzed in more detail in subsequent works.
	
	On the other hand, the former cognitive standards are completely suitable to be integrated by local governments into their prevention policies along with the corresponding bio-security measures. So, both dimensions, the biological and the cognitive, can work together for the improvement of human health globally.

	This is a small effort towards the development of more sophisticated techniques of a new kind of academic discipline that can be called "behavioural medicine".

	\section*{Acknowledgement}
		Danny A. J. Gomez-Ramirez would like to thank the Instituci\'on Universitaria Pascual Bravo and to the Instituci\'on Universitaria ITM (Instituto Tecnol\'ogico Metropolitano) and to Cognivision s.a.s. for all the support. He also wishes to thank William Restrepo, Elizabeth Yepes and Yury Baena for all their kindness and support. Yoe and Johana thank their daughter Salomé, Sofía y Sara for the joy and happines they bring.

	%\bibliographystyle{apa}
	%\bibliography{references}\printbibliography

\bibliographystyle{plain}

	\newpage

	\section*{Appendix} \label{ap:ap}
	%%%%%%%%%%%%%%%%%%%5
	%%%%%%%%%%%%%%%%%55
	%%%%%%%%%%%%%%%55
	\centering
	% include first image
	\includegraphics[width = 8cm]{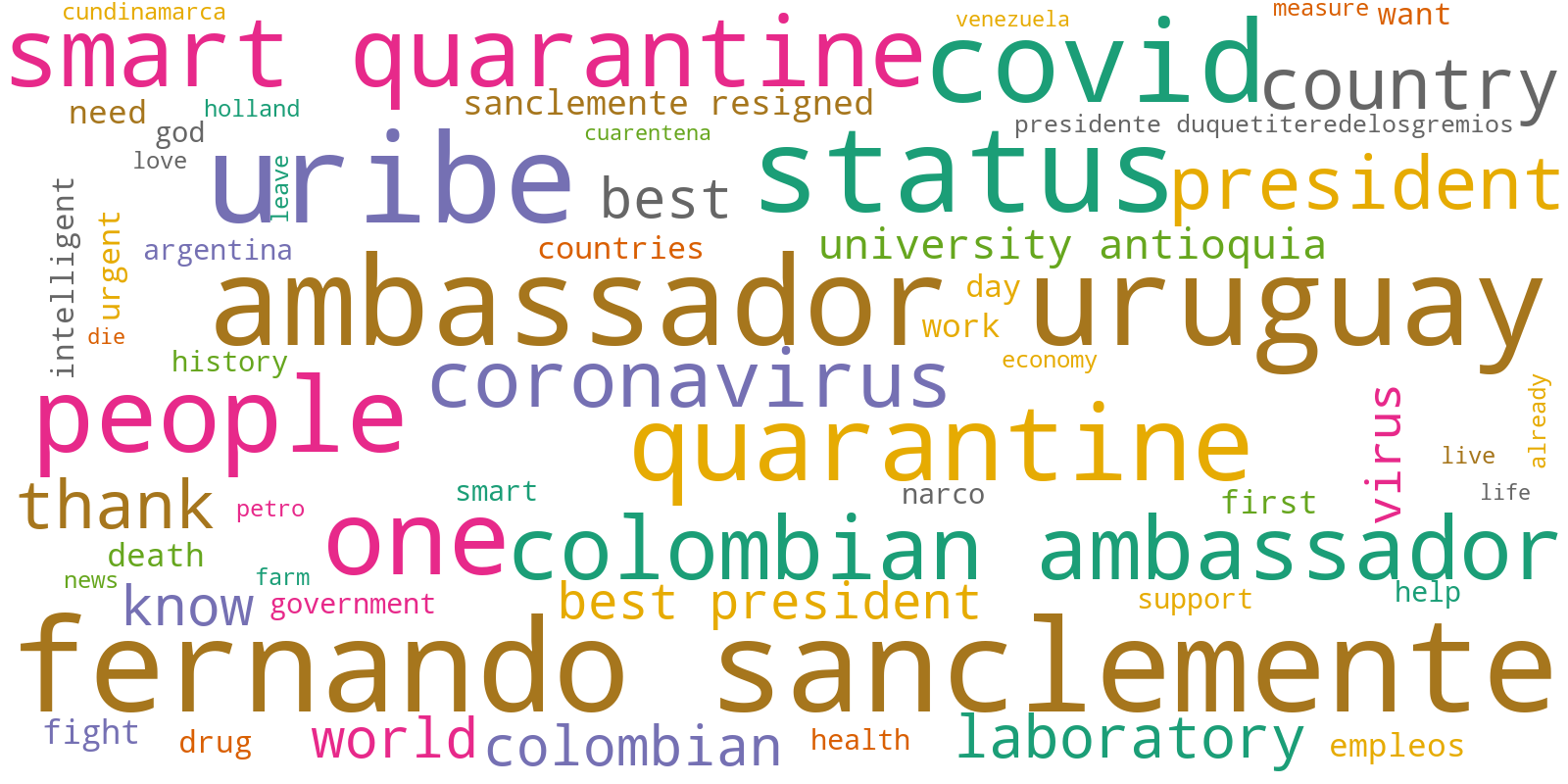}

	%%%%%%%%%%%%%%%%%%%5
	%%%%%%%%%%%%%%%%%55
	%%%%%%%%%%%%%%%55

	% include first image
	\includegraphics[width = 8cm]{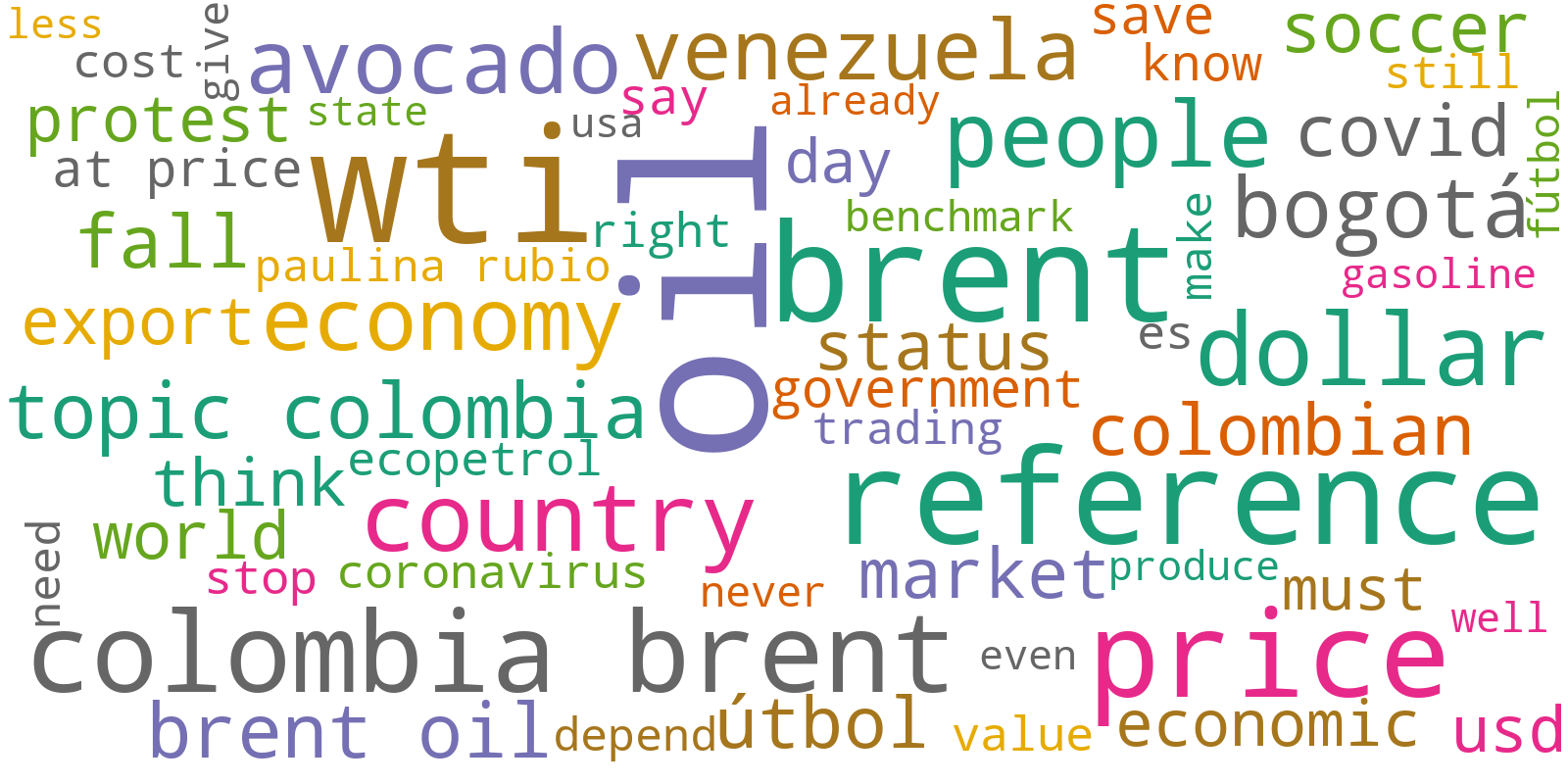}
	% include second image
	
	%%%%%%%%%%%%%%%%%%%5
	%%%%%%%%%%%%%%%%%55
	%%%%%%%%%%%%%%%55
	
	% include first image
	\includegraphics[width = 8cm]{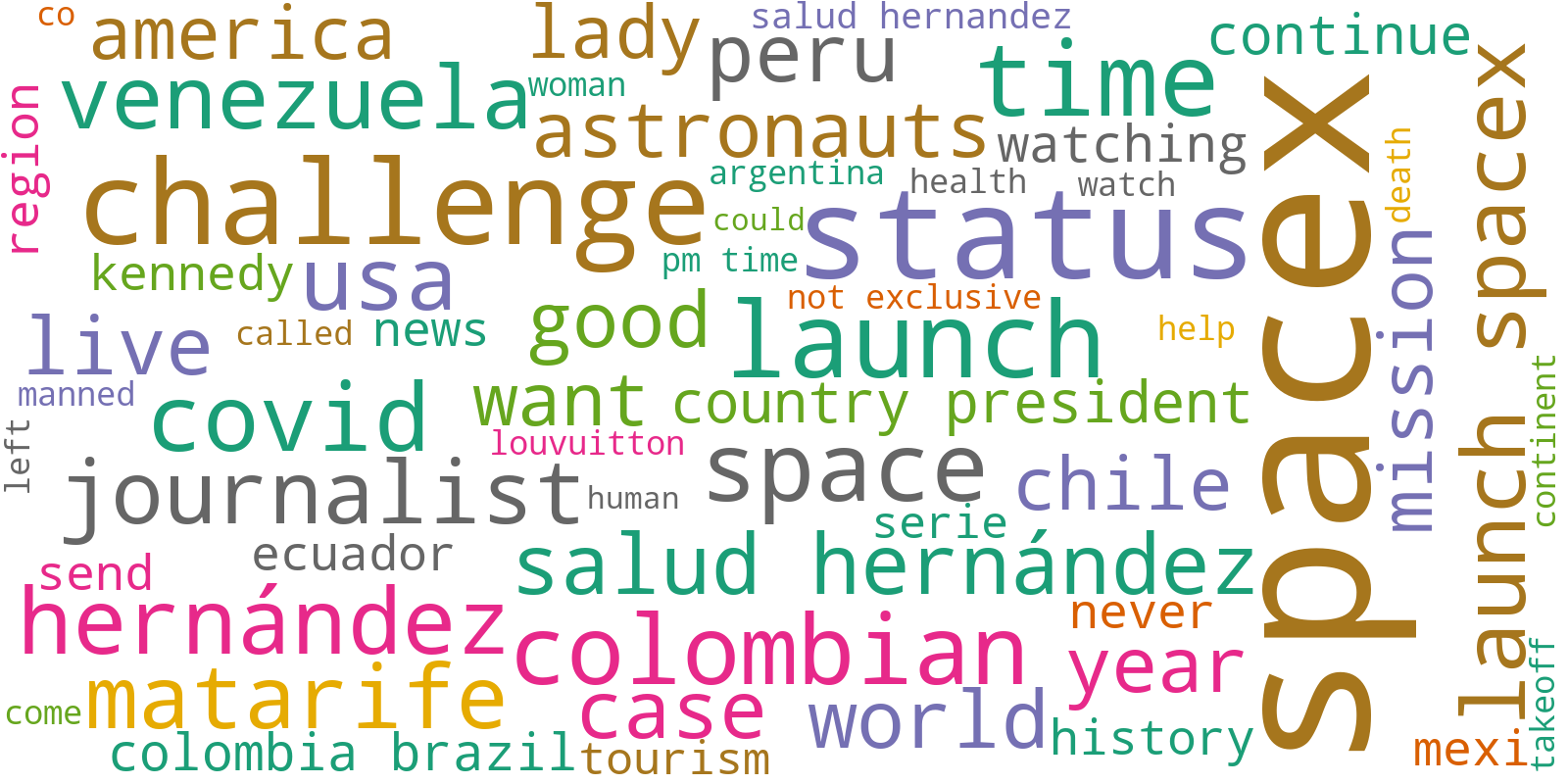}

	%%%%%%%%%%%%%%%%%%%5
	%%%%%%%%%%%%%%%%%55
	%%%%%%%%%%%%%%%55
	
	\includegraphics[width = 8cm]{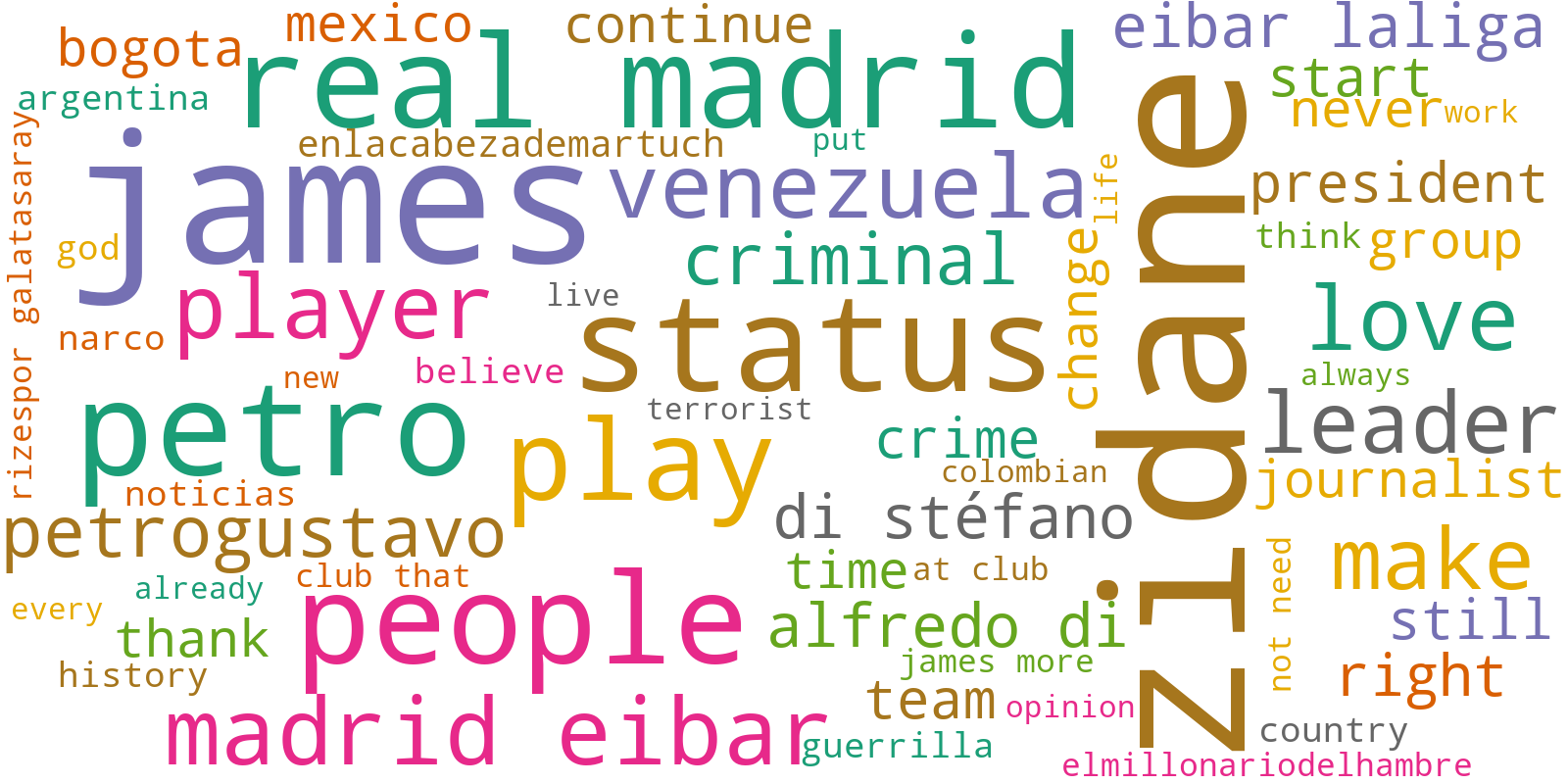}
	\begin{figure}[h]
		\begin{subfigure}{0.4\linewidth}
			% include second image
			\includegraphics[height=3.6cm,width=5cm]{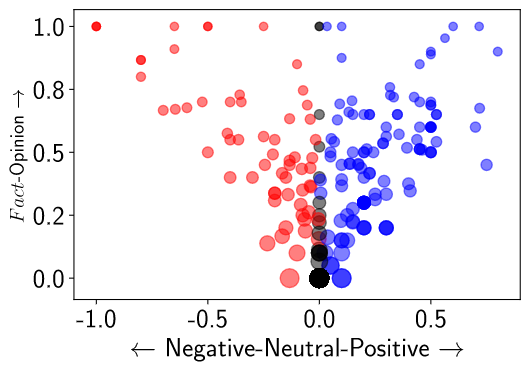}
		\end{subfigure}\hspace{1cm}
		\begin{subfigure}{0.5\linewidth}
			% include third image
			\includegraphics[height=4cm,width = 7cm]{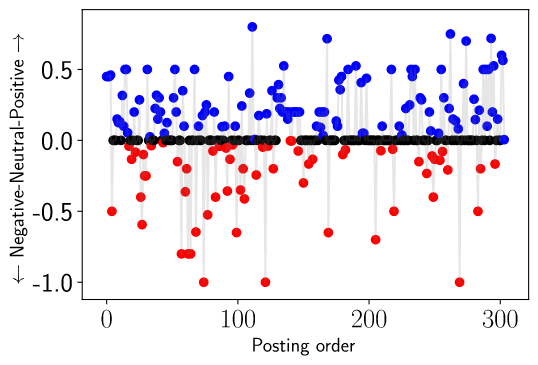}
		\end{subfigure}
		\caption{Sentiment analysis of June 14 and 15}
		\label{fig:day6}
	\end{figure}
	%%%%%%%%%%%%%%%%%%%
	%%%%%%%%%%%%%%%%%
	%%%%%%%%%%%%%%%
	
	\includegraphics[width = 8cm]{mar20_wc.png}

\end{document}